\documentclass[fleqn,10pt]{wlscirep}
\usepackage[utf8]{inputenc}
\usepackage[T1]{fontenc}
\usepackage{todonotes}
\usepackage{cleveref}
\usepackage{subcaption}

\title{Leveraging Uncertainty in Collective Opinion Dynamics with Heterogeneity}

\author[1,2,\dag]{Vito Mengers}
\author[1,3,\dag,*]{Mohsen Raoufi}
\author[1,2,\ddag]{Oliver Brock}
\author[1,4,\ddag]{Heiko Hamann}
\author[1,3,\ddag]{Pawel Romanczuk}
\affil[1]{Science of Intelligence, Research Cluster of Excellence, Berlin, Germany}
\affil[2]{Robotics and Biology Laboratory, Technische Universität Berlin, Berlin, Germany}
\affil[3]{Institute for Theoretical Biology, Department of Biology, Humboldt Universität zu Berlin, Berlin, Germany}
\affil[4]{Department of Computer and Information Science, University of Konstanz, Konstanz, Germany}

\affil[*]{mohsenraoufi@icloud.com}

\affil[$\dag$]{Equal Contribution}
\affil[$\ddag$]{Equal Supervision}

\begin{abstract}

Natural and artificial collectives exhibit heterogeneities across different dimensions, contributing to the complexity of their behavior. We investigate the effect of two such heterogeneities on collective opinion dynamics: heterogeneity of the quality of agents' prior information and of centrality in the network, i.e., the number of immediate neighbors. To study these heterogeneities, we not only consider them in our model, proposing a novel network generator with heterogeneous centrality, but also introduce uncertainty as an additional dimension. By quantifying the uncertainty of each agent, we provide a mechanism for agents to adaptively weigh their individual against social information. As uncertainties develop according to the interactions between agents, they capture information on heterogeneities. Therefore, uncertainty is a relevant additional observable in the study of complex collective opinion dynamics that we use to show the bidirectional relationship of heterogeneous centrality and information. Furthermore, we demonstrate that uncertainty-driven adaptive weighting leads to increased accuracy and speed of consensus, especially under heterogeneity, and provide guidelines for avoiding performance-decreasing errors in uncertainty modeling. These opportunities for improved performance and observability suggest the importance of uncertainty both for the study of natural and the design of artificial heterogeneous systems.

\end{abstract}
\begin{document}

\flushbottom
\maketitle
\thispagestyle{empty}

\section*{Introduction}

Individuals in collectives are exposed to a flow of incoming information from their neighbors, be it in a school of fish,~\cite{katz2011inferring} a network of sensors,~\cite{leonard2007collective, sirbu2017opinion, chen2023distributed} a swarm of robots,~\cite{atanasov2014information, hamann2018swarm, ebert2020bayes} or a group of humans.~\cite{moussaid2013social, pineda2015mass} Models of opinion dynamics not only suggest a mechanism for how individuals incorporate this information to update their internal opinions but also provide a way to design and engineer the equivalent artificial systems, e.g., distributed sensor networks~\cite{chen2023distributed} or multi-robot systems.~\cite{hamann2018opinion, Raoufi2023ICRA, leonard2023fast} Most of these models assume that all individuals are homogeneous to simplify the analysis of collective dynamics. However, this simplification does not reflect the extent of heterogeneity found in \emph{real} systems, natural~\cite{page2010diversity, apfelbaum2014rethinking, phillips2012delusions, ravary2007individual, d2017individual} or synthetic:\cite{raoufi_individuality_2023} The inter-individual variations of agents can manifest in multiple aspects, e.g., in behavioral traits,~\cite{bierbach2017behavioural} position in the network,~\cite{brass1984being, borgatti2009network} information access,~\cite{poel2022subcritical} %
or self-confidence.~\cite{kennedy2013overconfidence}

Heterogeneities change collective behavior in various ways, but to study their influence on opinion dynamics, we need to integrate them into the model. However, classical models of opinion dynamics treat heterogeneous sources of information, for example, different neighbors or environment information, equally.~\cite{kameda2022information} Thus, they do not provide a general mechanism for how agents should update their opinions in heterogeneous settings. 
To design such a mechanism, we turn to the field of robotics, where a similar problem occurs: Instead of heterogeneous agents in a collective system, a single robotic system already consists of heterogeneous components, such as different sensors, e.g., cameras, microphones, or force sensors, and functional units, e.g., perception, planning, or control. Since these heterogeneous components solve related problems, they need to share information.\cite{alatise2020review,eppner2016lessons,martin2022coupled} Therefore, each component must combine the information from heterogeneous sources, just as each agent does in a heterogeneous collective. In the robotic systems, we can achieve this by adding the dimension of uncertainty:~\cite{martin2022coupled} By quantifying the uncertainty of each component with probabilistic modeling,\cite{kalman1960new,julier1997non,thrun2005probabilistic} each component can combine the information of others according to their uncertainty. Because the uncertainties adaptively reflect the quality of information in the heterogeneous system, considering them allows each component to perform robustly under the noisy and ambiguous conditions of the real world.\cite{martin2022coupled,mengers2023combining}

In this paper, we apply the same principle of uncertainty-driven interaction to collective opinion dynamics models. We add uncertainty to the interactions between agents and thereby provide them with a mechanism to \emph{adaptively} weigh their available private and social information. The uncertainty of each agent dictates to what degree its opinion should be taken into account. Throughout the interactions between agents, their uncertainties change, leading to further adaptations of weights.
This sets our model apart from the classic {DeGroot} model,~\cite{degroot1974} where only the current opinion of each agent is tracked, and the weighting of different opinions remains static. While these static models work well in homogeneous settings, the missing adaptation is more evident in heterogeneous ones. Thus, other models have made weighting adaptive by employing different heuristics under specific assumptions.\cite{de2014learning,anderson2019recent,ding_consensus_2019} For example, in bounded confidence models\cite{hegselmann2002opinion} agents average the opinion of all other opinions that are within a \emph{confidence bound} around their own opinion, leading to adaptive but equally weighted interactions. These models allow studying heterogeneity with regard to their bound parameter,\cite{mirtabatabaei2012opinion,liang2013opinion} but are unable to represent many other types of heterogeneity, like of node degree.\cite{ding_consensus_2019}
Similar to our model, prior approaches have added the dimension of uncertainty to opinion dynamics to implement adaptive weights. These models have either focused on Bayesian\cite{ebert2020bayes} and non-Bayesian\cite{allahverdyan2014opinion,BASHARI2023theoretical} inference for discrete opinions, stochastic differential equations as opinion representation,\cite{de2016learning} or Gaussian message passing to infer the uncertainty of each agent as a full Bayesian posterior.\cite{PLARRE2004extended,chamley2013models} We similarly use Bayesian inference to estimate uncertainty over continuous heterogeneous opinions. But similar to the related robotics work,\cite{martin2022coupled} we employ and compare two \emph{simpler} approximations of an agent's uncertainty based only on its local neighborhood, because inferring a full posterior is an NP-hard problem.\cite{hazla_bayesian_2021,chickering_large-sample_nodate} We will show that the resulting adaptivity of our model leads to two advantages: increased performance compared to static models, in particular under heterogeneity, and robustness against modeling errors in most scenarios.

By introducing uncertainty to the interactions between agents, we not only gain adaptive weighting but, importantly, also an additional observable dimension. This dimension of uncertainty lets us study more complex opinion dynamics because it develops according to the \emph{heterogeneous} interactions between agents. We use this new observable to study two types of heterogeneity between agents of the collective: varying information quality and differing number of neighbors, i.e., node degree. These two dimensions often appear at the same time, as observed in natural systems. For instance, \emph{less central} fish on the periphery are \emph{more informed} about the surroundings of the collective.\cite{davidson2021collective,poel2022subcritical} We will show that the two dimensions of heterogeneity are coupled via uncertainty, meaning that varying one influences the other.

Heterogeneity of information quality means that some agents are more informed than others. Such situations can arise in collectives due to various reasons such as the inter-individual variations in perception, information gathering, access to information, or learning and experience.~\cite{jeanson2014interindividual, raoufi_individuality_2023} For instance, differences in information quality can be observed when fish on the periphery excel at detecting predators more effectively and rapidly than others.\cite{davidson2021collective,poel2022subcritical} These differences in information also lead to observable differences in behavior: Leaders within the collective typically exhibit greater certainty in their opinion,~\cite{hogg2001social} ants show differences in their exploratory behavior,~\cite{bonabeau1999swarm, d2017individual} and the waggle dance of bees with higher confidence in nest selection intensifies.~\cite{seeley1991collective} These disparities in information quality reflect on the uncertainty of agents, i.e., more informed agents are more certain.~\cite{kameda2022information} Therefore, we model such information heterogeneity in opinion dynamics by introducing diverse initial uncertainties among agents. We then investigate the influence of this initial heterogeneity on the subsequent opinion dynamics with uncertainty.

\begin{figure}[b]
    \centering
    \includegraphics[width=\textwidth]{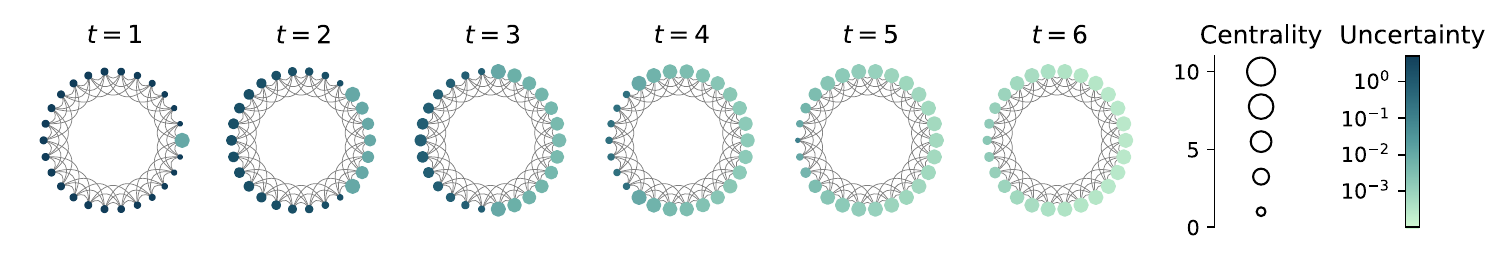}
    \caption{Centrality changes with uncertainty. By quantifying uncertainty in opinion dynamics, we can observe the complex development of centrality driven by relative differences in uncertainty. We show this for a $k$-regular homogeneous graph of $26$ agents with $k=6$. We distribute initial certainty heterogeneously by assigning higher certainty to only one agent. Following uncertainty-driven updating (BI-AI), agents with lower uncertainty relative to their neighbors gain higher in-degree centrality. %
    }
    \label{fig:kreg_uncert_central}
\end{figure}

Besides heterogeneity of information quality, we study heterogeneity of the number of neighbors an individual interacts with.
Agents with more neighbors are recognized to be more \emph{central}, as can be measured by in-degree centrality. %
Existing studies have demonstrated that these central nodes exert more influence over the collective opinion, and thus, draw the collective toward their opinion.~\cite{becker2017network} Given the bi-directionality of interactions between agents, having more neighbors also means that central nodes \emph{receive} more information.~\cite{stephenson1989rethinking}
To study these effects of the heterogeneity in centrality on collective opinion dynamics, we propose novel mechanisms for generating random heterogeneous networks. Compared to previous works,~\cite{erdds1959random, albert2002statistical} our primary emphasis in network generation is on maintaining constant network connectivity, i.e., mean degree, while modulating the variance of the degree distribution, comparable to the Watts-Strogatz approach starting from a regular network.~\cite{watts1998collective} This approach enables us to produce a range of networks with various centralization heterogeneity while keeping the spectral properties of the network constant. We will show how varying this centralization influences collective opinion dynamics, with and without uncertainty.

The remainder of this paper is organized as follows: We first analyze the general role of uncertainty in collective opinion dynamics and show the impact of centrality heterogeneity on certainty, and the other way around. Next, we compare different opinion updating mechanisms concerning their performance in various heterogeneous settings. Then we evaluate two different approaches to approximate uncertainty in terms of robustness against modeling errors. After discussing the results, we describe the methods and metrics used throughout the paper.

\section*{Results}
\subsection*{The Role of Uncertainty in Heterogeneous Collective Opinion Dynamics}

Using uncertainty in the collective opinion dynamics enables agents to reflect on the quality of information they share and to weigh the received opinions of their peers accordingly. 
This dimension of uncertainty has its own dynamics over time, which is a function of various factors, such as initial uncertainty distribution and the centrality of each node. Therefore, uncertainty contains information about different types of heterogeneity that affect collective information processing. 

In order to explore the impact of accounting for uncertainty on collective dynamics with heterogeneity, we compare four different models: two uncertainty-driven models with different uncertainty approximations and two naive averaging models with different weighting schemes. The uncertainty-based models both update opinion and uncertainty of each agent using Bayesian inference but differ in the assumption used to derive the updated uncertainty: either assuming independence of opinions (Bayesian inference assuming independence, BI-AI) or more conservatively assuming their unknown correlation (Bayesian inference assuming unknown dependence, BI-UD). Both of the naive methods on the other hand average the opinions of all neighbors and then combine them with the agent's current opinion following a self-weight parameter in classical DeGroot fashion. However, we set this self-weight differently for the different methods: either we set them according to the number of local neighbors to ensure fair averaging of individual and social information (naive averaging with locally equal weights, NA-LEW) or we assign a universal self-weight across all agents, which is optimized offline for performance (naive averaging with optimal universal weight, NA-OUW). For more details on their implementation, see the Methods section.

\begin{figure}[t!]
    \centering
    \includegraphics[width=\textwidth]{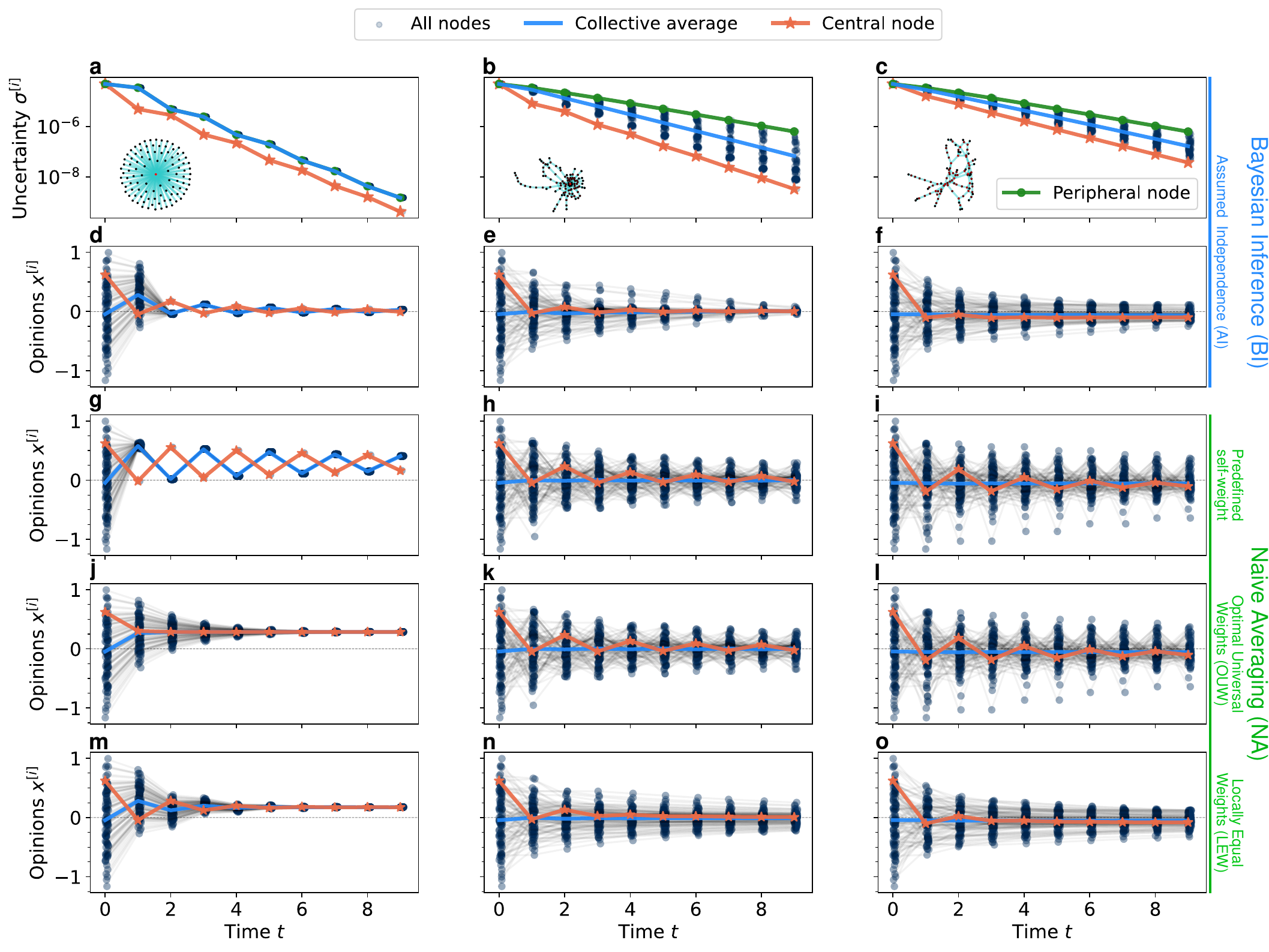}
    \caption{Centrality determines the rate of uncertainty reduction of nodes (\textbf{a}-\textbf{c}), while uncertainty provides an adaptive weighting mechanism (as in \textbf{d}-\textbf{f}). This adaptivity alleviates the negative effects observed in heterogeneous networks, such as central nodes pulling the collective opinion. This pulling effect can be seen in \textbf{g}, \textbf{j}, and \textbf{m}, with the collective consensus leaning toward the central node, as opposed to the neutral consensus in \textbf{d}.
    The first row (\textbf{a}-\textbf{c}) shows how the uncertainty of the most central node compares to that of the other nodes in a Bayesian approach (BI-AI) with  a homogeneous initial distribution of uncertainty. The following rows show the time-development of opinions of all nodes, of the central node, and the average opinion of the collective, for different opinion updating mechanisms.
    All settings share the same initial opinions, each column corresponds to a different network centrality, as shown in the first row. }
    \label{fig:overtime}
\end{figure}

\begin{figure}[t!]
    \centering
    \includegraphics[width=\textwidth]{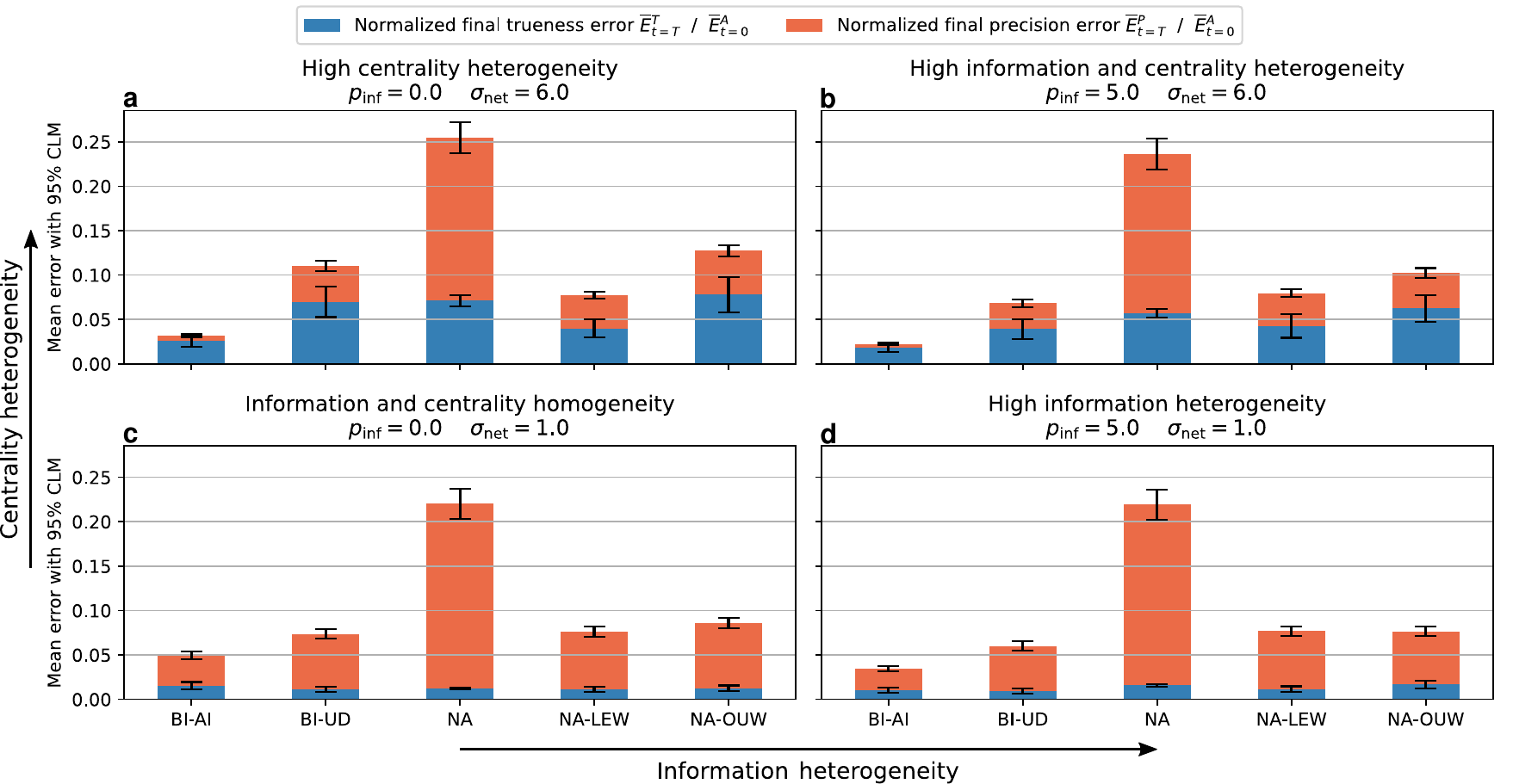}
    \caption{Quantifying uncertainty enables agents to outperform the naive averaging methods, especially in heterogeneous settings (\textbf{b}). The mean performance of naive averaging for different self-weights (NA) has low precision, having locally equal weights (NA-LEW) enhances this precision, similar to choosing an optimal universal weight (NA-OUW) for all agents. By assuming independence (BI-AI) agents capture more information about the heterogeneity in the network, leading to overall lower errors under heterogeneity, but higher bias (trueness error) in homogeneous conditions (\textbf{c}).
    }
    \label{fig:method_comp}
\end{figure}

\begin{figure}[b!]
    \centering
    \includegraphics[width=\textwidth]{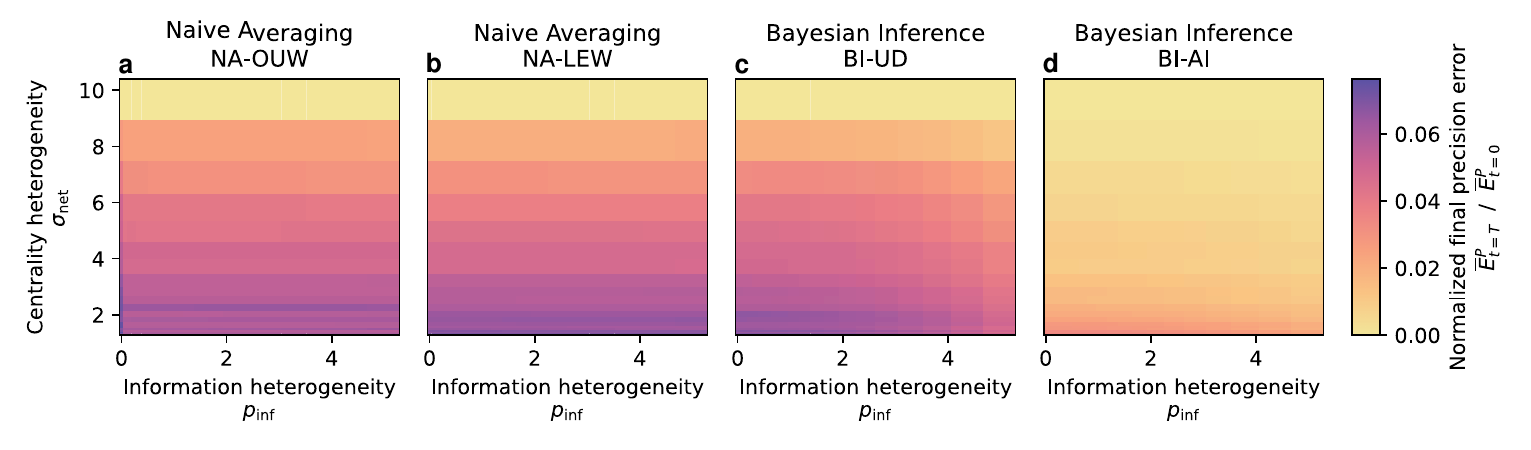}
    \caption{Heterogeneity speeds up the collective consensus. 
    While Bayesian inference methods (\textbf{c}, \textbf{d}) leverage the heterogeneity of information \emph{and} centrality, naive averaging methods (\textbf{a}, \textbf{b}) lack a mechanism to take advantage of information heterogeneity. The independence assumption accelerates the consensus of the Bayesian approach even further (BI-AI, \textbf{d}).
    We show this by evaluating the precision error for different methods under two different dimensions of heterogeneity. The final precision error normalized with respect to the initial one indicates the speed of convergence, given a fixed amount of time. 
    }
    \label{fig:het_het_comp}
\end{figure}

Following uncertainty-driven mechanisms, agents weigh their incoming information adaptively. This directly reflects upon the overall collective, leading to more complex opinion dynamics, that adapt according to heterogeneity, as shown in \Cref{fig:kreg_uncert_central,fig:overtime}. We can see that this adaptivity leads to differences in collective performance compared to the static naive approaches, in particular under increasing heterogeneity. In \Cref{fig:method_comp,fig:het_het_comp}, we can see how static methods are practically unchanged by heterogeneity in information, while the adaptivity of the uncertainty-driven mechanisms lets them harness the available information. Heterogeneity in centrality on the other hand influences all methods, but in particular BI-AI can take advantage of this heterogeneity by adapting accordingly. 

In the following sections, we first study how the two heterogeneities, of initial information quality and of the centrality of the nodes in the network, act on the collective opinion dynamics with uncertainty. In particular, we will describe how centrality and uncertainty influence each other under Bayesian inference with assumed independence (BI-AI). After this analysis of uncertainty-driven dynamics, we will analyze how their adaptivity influences performance compared to static methods.

\subsubsection*{Centrality of Nodes in the Network Influences Uncertainty}

The position of a node in the network determines the flow of information it receives. In particular, nodes with a higher degree centrality are connected to more agents and thus have access to more social information, increasing their certainty over time. To show this effect, we consider agents with the same initial uncertainty but heterogeneous centrality within the network. 

The results in \Cref{fig:overtime}a show that the central node in a fully centralized star network achieves more certainty over time, compared to the peripheral nodes, despite sharing the same initial certainty. This is because all the peripheral nodes share their opinion with the central node, while they only receive one opinion from it in return. The results for the less centralized networks (\Cref{fig:overtime}b-c) suggest a similar but less pronounced effect and thus confirm how network position influences the uncertainty of nodes: \emph{Centrality in the network determines the rate of change of uncertainty.}

\subsubsection*{Uncertainty Influences Centrality} 
The relation between uncertainty and centrality is bi-directional, meaning that, under specific conditions, more informed nodes influence the collective opinion dynamic as if they would be more central than others, despite having the %
same position on the graph. To show this, we consider a homogeneous $k$-regular network (where each node has exactly $k$~neighbors) and distribute uncertainty heterogeneously: One of the agents starts with much lower uncertainty than all others. 
Since the weights of the links are determined based on the \emph{differences} in uncertainty of agents, the \emph{weighted} network is heterogeneous, despite having a regular, homogeneous graph structure. Concretely, the in-degree centrality of nodes is heterogeneous as it is determined by the relative uncertainties. For example, the most certain agent at time $t=1$ in~\Cref{fig:kreg_uncert_central} is the most central node. 
At the same time, nodes that are immediately adjacent to the most certain node are less central, despite having the same uncertainty with the other neighbors.

Over the next processing steps, the neighbors of the most certain nodes decrease their uncertainty quickly. This leads to the centrality shifting away from the initially certain node to the nodes that are at each step on the border between certain and uncertain agents. The uncertain neighbors on the border are the least central nodes at the same time, as explained before. This demonstrates how uncertainty influences centrality in collective opinion dynamics: \emph{Relative differences in uncertainty between connected agents lead to differences in centrality}.

\subsection*{Performance of Uncertainty-Driven Opinion Models}

\subsubsection*{Uncertainty-Driven Opinion Dynamics Are Faster}

Agents that have access to more information, be it due to more neighbors or better initial information, have lower uncertainty and thus pull their neighborhoods more quickly toward a consensus. Similarly, their neighbors become more certain in the process and can thus convince their respective neighbors.
As shown in \Cref{fig:method_comp}, the accelerated local consensus leads to higher global precision, which implies a faster convergence 
given the fixed duration we considered.
It is driven by the adaptive centrality in the collective caused by differing uncertainties, as described before. Thus, under greater differences among agents caused by higher heterogeneity in uncertainty or centrality, the convergence becomes even faster, shown by reduced final precision error under increasing heterogeneities in~\Cref{fig:het_het_comp}. This demonstrates a key benefit of considering uncertainty in heterogeneous collective opinion dynamics:  \emph{Because uncertainty captures heterogeneity within the collective, its consideration leads to faster convergence.}

If we compare the uncertainty-driven with naive methods in \Cref{fig:method_comp}, we find that the uncertainty-driven methods (BI-AI, BI-UD) perform similarly to the locally equal (NA-LEW) and optimally universally (NA-OUW) weighted naive methods in homogeneous settings, while outperforming them in heterogeneous settings. Comparatively, the mean across all possible values of a universal weight (NA) has much higher errors in any setting, showing the importance of setting weights correctly if they remain static. The adaptive weighting of the uncertainty-driven methods on the other hand does not require weighting parameters but requires initial uncertainty quantification and inference assumptions. To evaluate how these influence the performance of uncertainty-driven opinion dynamics, we will now first compare the effect of the independence assumption as compared to Bayesian inference under unknown dependence, and then evaluate the impact of errors in quantifying uncertainty.

\subsubsection*{The Role of the Independence Assumption}

Quantifying the uncertainty of each agent with Bayesian inference requires information about the correlation of the opinions of neighboring agents. Determining this correlation exactly requires recording all opinion exchanges in the network, which, as we mentioned, is an NP-hard problem. We instead use two local approximations of uncertainty with different assumptions, assuming independence of all opinions (BI-AI) or their unknown dependence (BI-UD). For more details, see the Methods section. If we assume independence of opinions, the uncertainty of agents with more neighbors decreases faster, because they have access to more \emph{independent} information. If we assume unknown dependence, we have to conservatively approximate the updated uncertainty to be not smaller than the smallest uncertainty in the neighborhood, meaning that centrality does not directly influence the uncertainty. Therefore, the bidirectional influence of centrality and uncertainty only takes place under the independence assumption. This influence allows BI-AI to outperform other methods under heterogeneity, as can be seen in ~\Cref{fig:method_comp}. Moreover, the monotonic decrease of uncertainty under the independence assumption leads to faster convergence.

\begin{figure}[b!]
    \centering
    \includegraphics[width=\linewidth]{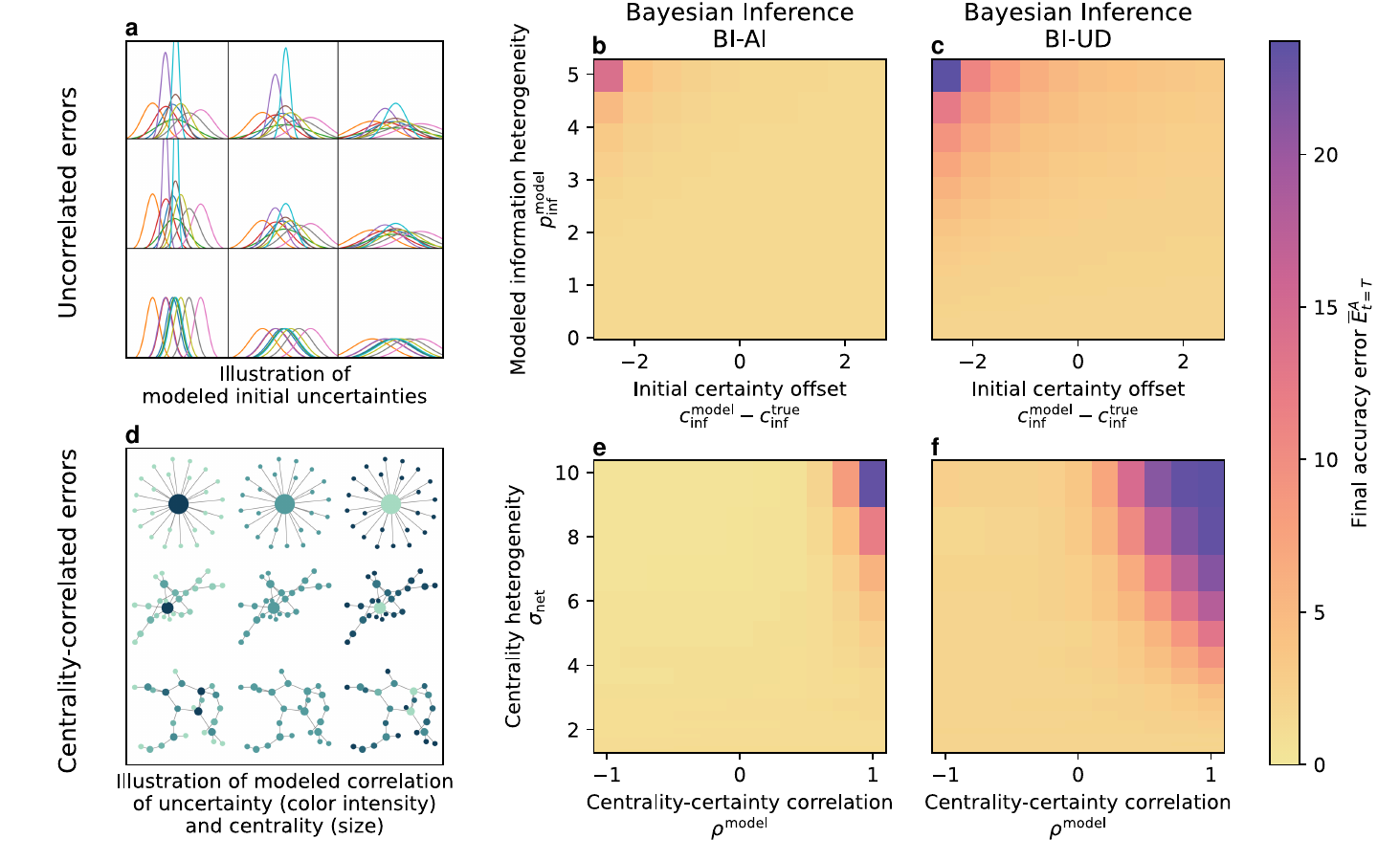}%
    \caption{Uncertainty-driven methods are robust against modeling errors unless agents are highly over-confident, in particular if they are central.
    We show this by evaluating performance under varying uncorrelated systematic errors (illustrated in \textbf{a}) and varying centrality-correlated errors (illustrated in \textbf{d}) for different Bayesian mechanisms.
    The performance, measured as accuracy error, only decreases for uncorrelated errors when the information heterogeneity is high and agents are over-confident about their uncertainty (top-left corner in \textbf{b} and \textbf{c}). Whereas for centrality-correlated errors, the combination of heterogeneous networks and over-confident central nodes (top-right corner in \textbf{e} and \textbf{f}) is detrimental to the collective performance.
    Initial opinions are drawn from a Gaussian distribution, but uncertainties are assigned according to the modeling error.
    }
    \label{fig:uncertainty_model}
\end{figure}

However, the opinions of neighbors are \emph{not} independent, and we see in \Cref{fig:method_comp} that assuming independence (BI-AI) leads to higher bias in homogeneous settings where leveraging the heterogeneity in the system is not possible. For BI-UD, we on the other hand obtain performance similar to optimally weighted naive averaging for different networks. Under heterogeneity of information, both bias (trueness error) and convergence speed (related to precision error) improve, as the high-value information of some agents can be used to improve collective performance. Thus, the assumption on correlation leads to different strengths: \emph{Assuming independence allows leveraging heterogeneity, but assuming unknown dependence ensures low bias.}

\subsubsection*{The Effect of Errors in Quantifying Uncertainty}
Adding uncertainty as a variable to measure the quality of information also brings a new challenge for agents: Quantifying uncertainty can be prone to errors. These errors alter the opinion dynamics in collectives, especially when they exhibit high heterogeneities. 
Previously, we assumed that there was no mismatch between the actual quality of initial information agents received, and what they assumed. Here, we intentionally model errors in quantifying uncertainty, to investigate how they influence collective opinion dynamics.

If we consider uncorrelated errors, we have a spectrum of collectives where agents are on average either over-confident or under-confident (see illustration in \Cref{fig:uncertainty_model}a). Our results in \Cref{fig:uncertainty_model}b-c show that these errors do not affect the collective performance significantly, unless for the over-confident collectives with heterogeneous information. Under low heterogeneity or systematic under-confidence, the accuracy of consensus is not impacted as significantly.

Assuming errors that are correlated with the centrality of each agent, as in \Cref{fig:uncertainty_model}d-f, we can observe that errors only rise strongly if centrality and certainty are \emph{positively} correlated. In this case, there exist central nodes that are extremely over-confident and can sway the whole collective opinion. Thus, leading to high errors if their opinion is far from the initial collective average. This effect is stronger for centralized collectives. 

For both uncorrelated and correlated errors, inference assuming unknown dependence (BI-UD) is more vulnerable in the respective areas of increasing error than under independence assumption (BI-AI). This is because BI-AI can alleviate the effect of errors due to its ability to leverage heterogeneity, as described in the previous section. Overall, the areas of increasing errors are similar for both methods, with most of the modeled error configurations leading to only small performance decreases, suggesting that 
\emph{uncertainty-driven methods are robust to errors in quantifying uncertainty as long as uncertainty estimates are under-confident, in particular for central nodes}.

\section*{Discussion}
Heterogeneities are common in both natural and artificial systems. 
In collective opinion dynamics, they result in more complex behavior, as we have shown.
For instance, informed nodes can be more influential on the collective consensus. 
Observing such behavior in models can only be possible if we assume that agents' information is heterogeneous, for example, due to having different amounts of prior information, or due to obtaining more information over time. 
But assuming such heterogeneity without accounting for it in opinion updating mechanisms can often lead to undesirable effects, such as central nodes pulling the whole collective toward their opinion.
We can alleviate such effects by quantifying the uncertainties of agents.
These uncertainties provide a mechanism for agents to adaptively weigh their individual against social information while developing according to the interactions between agents. Therefore, this additional dimension of uncertainty can capture heterogeneity in collectives, as we have shown for heterogeneity in centrality and information. This way, uncertainty-driven opinion dynamics do not just alleviate undesirable effects but also allow leveraging heterogeneity to obtain faster consensus. 

Compared to the naive averaging methods, uncertainty-based models have higher accuracy and accelerated consensus, especially in more heterogeneous settings. As we have demonstrated, these benefits can be achieved even if we quantify uncertainty with only local approximations, instead of tracking the full history of uncertainty values. 
Moreover, this way of approximating uncertainty performs robustly against modeling errors, unless agents are systematically over-confident, in particular for central nodes.
Due to the benefits and robustness of uncertainty models, they should be considered when designing artificial systems with opinion dynamics or collective estimation, such as sensor networks or multi-robot systems. Considering uncertainty comes at the cost of increased computation and communication, which nevertheless are still of the same order (e.g., $\mathcal{O}{(N \cdot k)}$, for $k$-regular networks of $N$ agents) as naive averaging methods when uncertainty is only approximated locally.

Besides their advantage in terms of performance, the dimension of uncertainty offers a new observable with its own dynamic.
Tracking uncertainty thus opens up new opportunities to study collective opinion dynamics, in particular under heterogeneity. In this paper, we used uncertainty to present the relationship of heterogeneity in centrality and certainty, highlighting their bidirectional correlation. 
Understanding this relationship further will shed light on the collective information processing in natural systems, as information in these systems is often distributed unevenly. 
In fish swarms, for instance, only the peripheral fish can notice an appearing predator.\cite{davidson2021collective,poel2022subcritical}

However, considering uncertainty as an additional observable is not limited to the here-considered heterogeneities. 
As we described before, real systems are heterogeneous along many axes. Heterogeneities caused by the dynamic nature of the environment, or a dynamic interaction network, for example, can also influence the uncertainty of agents. 
In such cases, extensions of the model for the time-development of uncertainty of each agent might be required. 
More natural scenarios might also be better addressed by modeling indirect communication of uncertainty, for example, by observing the activity of other agents, or by estimating their uncertainty.
Since more certain and central agents can sway the collective easily, incentives to malicious behavior might also be observed. Consequently, scenarios where malicious agents communicate wrong uncertainty to influence the collective more severely, suggest a framework to investigate more complex collective opinion dynamics.

As we described in the beginning, our uncertainty-driven approach for collective opinion dynamics under heterogeneity mirrors an approach from robotics, which was designed for the interaction of multiple heterogeneous components.\cite{martin2022coupled} But instead of complex information exchanges between different components within one robot, here we investigated the
exchange of scalar opinions between agents within a collective. As we have demonstrated, we can apply models following the same principles to these two problem domains, because they share a similar structure. This suggests relevant connections of the fields of robotics and collective dynamics beyond their traditional boundaries, that meet mostly in collective robot systems~\cite{leonard2007collective} and collective estimation.~\cite{Raoufi2023ICRA} Given their similar structure, insights from each field might be relevant to the other. While we show how taking an idea from robotics to the field of collective opinion dynamics proves fruitful, the opposite direction might also be interesting. 
For instance, compared to robotic systems with exchanges of high-dimensional data in varying representations, observing the exchange of scalar opinions in collectives simplifies the identification of global effects.
These effects such as the bidirectional influence of uncertainty in centrality might at the same time be relevant to the design of robotic systems because they influence the global performance. Thus, we anticipate that further integration of these fields will be fruitful. %
But this is not limited to robotics or collective dynamics since other disciplines also face problems of similar structure. In vision science for example, models following the same principles as our opinion dynamics model have been shown to aid in the study of how humans perceive visual illusions,~\cite{godinez2023probing} suggesting a shared underlying problem structure.

\section*{Methods}

We investigate the effect of heterogeneities on collective opinion dynamics using different models to update individual opinions. In the following, we will describe how we generate collectives with two types of heterogeneity: heterogeneity of centrality within the network and heterogeneity of the initial information quality of different individuals. We then detail the methods we used in our experiments to model the opinion updating of agents and the metrics used to evaluate the effect of heterogeneities on them.

\begin{figure}[b!]
\captionsetup[subfigure]{labelformat=empty}
\centering
    \subcaptionbox{}{\includegraphics[width=0.48 \linewidth]{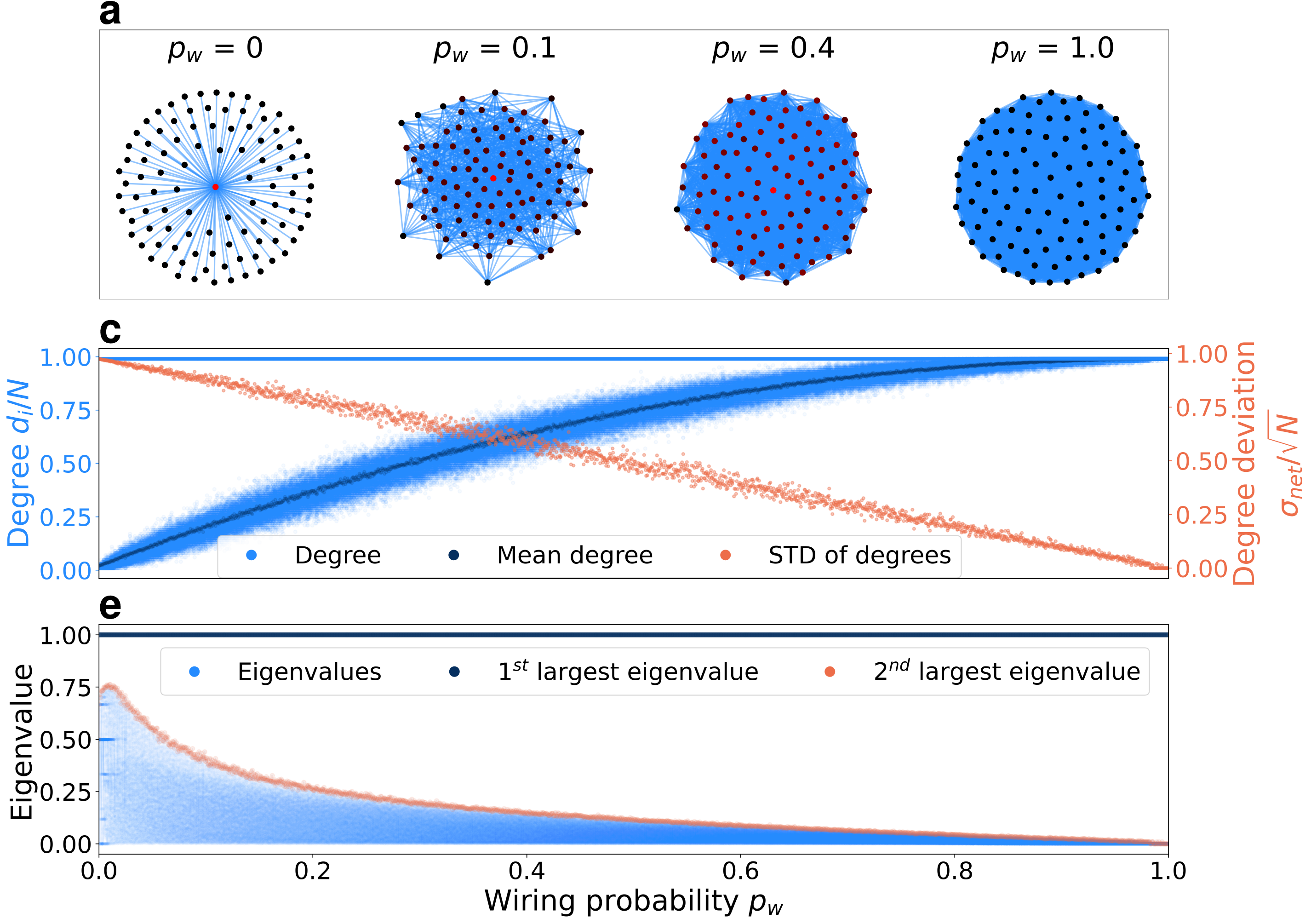}}%
    \hfill
    \subcaptionbox{}{\includegraphics[width=0.48 \linewidth]{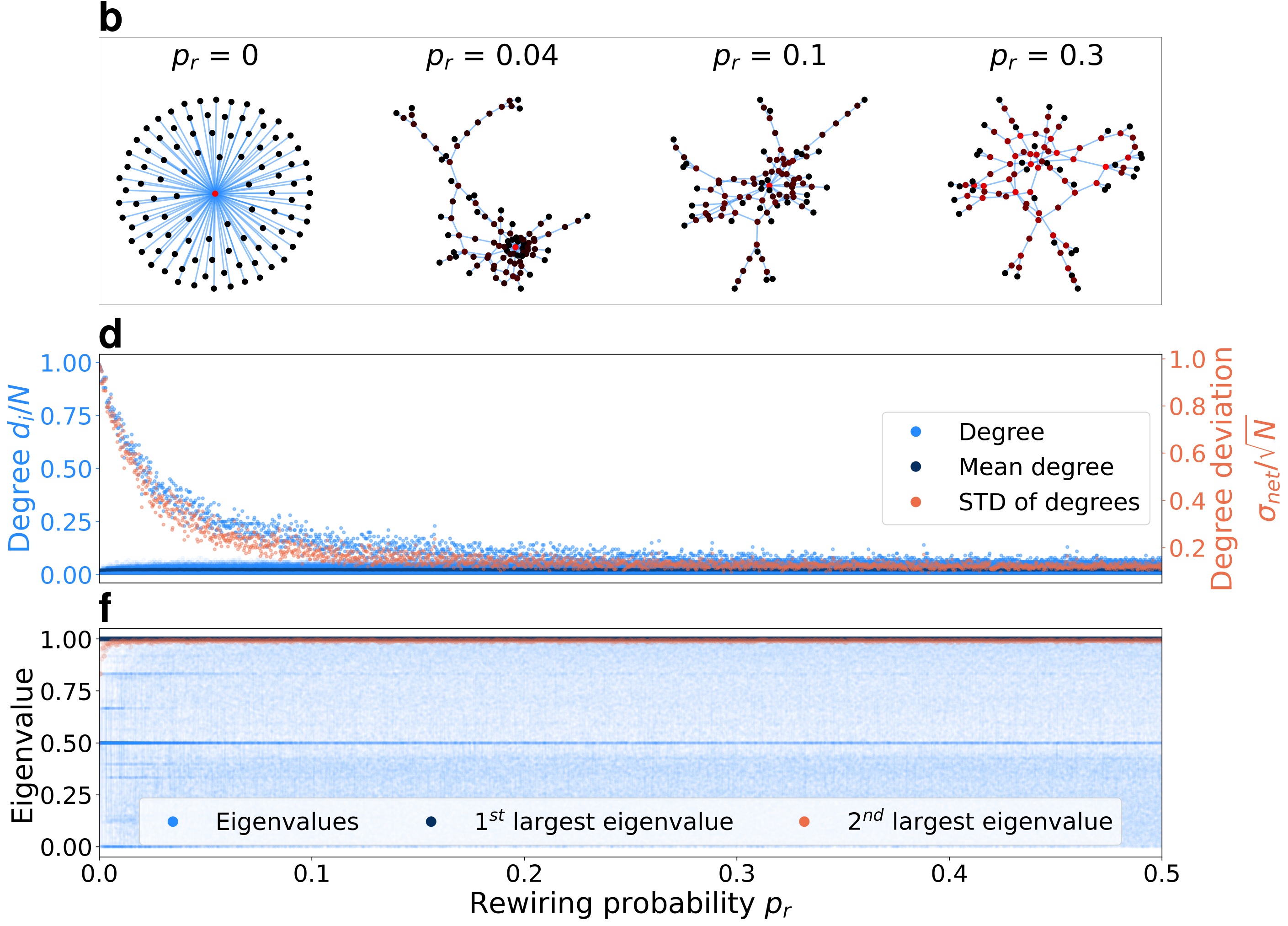}}%
\caption{Comparison of different models for generating heterogeneous graphs. On the left, we generate new graphs by adding more links to the star graph, and on the right by adding \emph{and} replacing the existing links. We show four different network samples generated by each model for different (re-)wiring probabilities in \textbf{a} and \textbf{b}, the degree-centrality of each node is color-coded by its red intensity. In \textbf{c} and \textbf{d}, we illustrate the degree distribution for different (re-)wiring probabilities, with the mean and standard deviation of all nodes. We quantify the spectral properties of the network with the left eigenvalues of the row-normalized adjacency matrix shown in \textbf{f} and \textbf{e}. The largest eigenvalue remains one proving a connected network, while the second largest eigenvalue changes by the mean degree.}
    \label{fig:net_gen_diff_methods}
\end{figure}

\subsection*{Heterogeneous Network Generation}
We define the network heterogeneity with regard to the variance of degree distribution.~\cite{snijders1981degree} We measure this by calculating the standard deviation $\sigma_\text{net}(d)$ of a network with size $N$ as explained later in the Metrics section. To study the effect of network heterogeneity, a spectrum of random graphs with varying heterogeneity is needed. One condition, however, is that all these graphs must be connected, meaning that there is a path between any two random nodes. In the following, we compare two different types of graph generator models. We use the term ``graph'' to refer to the structure of the \textit{unweighted} network, while the links of the ``network'' can be weighted.
\subsubsection*{By Adding More Links}
We assume the star graph $G^*$ to be at the very centralized extreme of the spectrum. By adding a link between a random pair of nodes, the network becomes incrementally more connected and shifts toward the decentralized extreme, which is an all-to-all network. The probability of selecting a pair of nodes is determined by the wiring parameter $p_\text{w}$. A few examples of the random graphs generated by this model are illustrated in~\Cref{fig:net_gen_diff_methods}a. The standard deviation of the degree distribution decreases linearly by the wiring probability $p_\text{w}$, suggesting a complete transition from a fully centralized heterogeneous graph to a fully decentralized homogeneous graph. In this model, adding new links to the graph increases the mean degree (\Cref{fig:net_gen_diff_methods}c). Consequently, the spectral properties of graphs generated in this model for different wiring probabilities are significantly different (see \textbf{e} in~\Cref{fig:net_gen_diff_methods}). Therefore, a comparison of speed or precision for different graphs is unfair, due to the influences caused by the increased connectivity in the graph. In our previous work,~\cite{Raoufi2023ICRA} we showed how the spectral properties of graphs, and in particular, their second-largest eigenvalue, affect the speed and precision of convergence in collective estimation. In the following, we propose an alternative approach that keeps the mean degree constant while varying the heterogeneity.
\subsubsection*{By Replacing Existing Links}
The first step is to generate a random, connected graph $G_1(\text{N},p_\text{r})$ by adding links to the star graph $G^*$. However, to ensure a fixed mean degree, we need further steps compared to the previous model. Assuming that we added $m$ new links in the first step, we need to remove $m$ random links among the existing links to return the mean degree to that of the $G^*$. This process results in another random graph $G_2(\text{N},p_\text{r})$, which might be disconnected due to the removal of the connecting links. A disconnected graph violates the condition we assumed earlier. 
In the next step, we make the network connected by adding links between two disconnected components. To do so, we pick two random nodes from two random components and add a link between them. We repeat this process until there are no more disconnected components left. The final \textit{connected} random graph $G_3(\text{N},p_\text{r})$ has a mean degree approximately the same as a star graph (see~\Cref{fig:net_gen_diff_methods}d). The degree distribution changes by varying $p_\text{r}$, particularly, the standard deviation, indicating centrality heterogeneity, decreases while the mean degree is kept almost constant. The reduction of standard deviation reaches a stationary state at around $p_{\text{r}}=0.3$. Therefore, we limited the range of $p_\text{r}$ to $[0,0.3]$ in our experiments reported in the Results section. 
The networks generated by this method show little variation in the eigenvalues of their row-normalized adjacency matrix (~\Cref{fig:net_gen_diff_methods}f). The largest eigenvalue remains at one, proving the graph being connected, and the second-largest eigenvalue remains fixed, indicating little change in the mean degree of graphs.

\subsection*{Heterogeneous Distribution of Initial Information}

All agents within the collective receive an initial opinion on the relevant state. We assume this initial opinion ${x_0}^{[i]}$ of the agent $i$ reflects a neutral collective average opinion $x_\mathrm{neutral}$ but with added noise $\eta^{[i]}$. This noise follows a zero-mean Gaussian with a variance of ${\sigma_0^2}^{[i]}$. 
It is important to distinguish between the diversity of initial opinions (stemming from $\eta^{[i]}$) and the heterogeneity of their quality.
Initial information quality becomes heterogeneous when the variance ${\sigma_0^2}^{[i]}$ of the added noise varies between agents. In real collectives, this might be either due to some agents being experts, due to only some agents being able to sense the relevant state of the environment, or due to previous exchange of opinions. We model such heterogeneity by sampling the standard deviation ${\sigma_0}^{[i]}$ for each agent $i$ from a uniform distribution centered around $c_\mathrm{inf}$ with a width of $p_\mathrm{inf}$.

\begin{gather}\label{eq:information_distr}
  {x_0}^{[i]} = x_\mathrm{neutral} + \eta^{[i]} \qquad \qquad \eta^{[i]} \sim \mathcal{N}(0,\,{\sigma_0^{2}}^{[i]}) \qquad \qquad {\sigma_0}^{[i]} \sim U(c_\mathrm{inf}-\frac{p_\mathrm{inf}}{2},c_\mathrm{inf}+\frac{p_\mathrm{inf}}{2})
\end{gather}

For our experiments, we choose $c_\mathrm{inf}=5$ and vary $p_\mathrm{inf}$ between $0$ and $5$ to ensure only positive values for variance. This way we generate collectives with varying heterogeneity of initial information, as shown in the middle column of \Cref{fig:het_inf_gen} for 25 agents. We will later explain how different methods that start with these initial opinions update them over multiple steps for the collective to converge.

\subsubsection*{Distribution of Information to Investigate Errors in Uncertainty Quantification}
To investigate the impact of errors in quantifying uncertainty, we distribute information homogeneously but assign uncertainties to agents as if we had distributed it heterogeneously. We compare two different types of such errors, uncorrelated errors of the uncertainty and errors that are correlated with the centrality of agents.

For the uncorrelated case, we distribute information according to \Cref{eq:information_distr} with $c_\mathrm{inf}^\mathrm{true}=5$ and $p_\mathrm{inf}^\mathrm{true}=0$ but distribute the initial uncertainty following the same equation but with $c_\mathrm{inf}^\mathrm{model}$ between $2.5$ and $7.5$ and $p_\mathrm{inf}^\mathrm{model}$ between $0$ and $5$. This leads to collectives with errors in the amount of heterogeneity (varying $p_\mathrm{inf}^\mathrm{model}$) and errors that on average resemble under-confident or over-confident agents (according to the difference between $c_\mathrm{inf}^\mathrm{true}$ and $c_\mathrm{inf}^\mathrm{model}$). For the correlated case on the other hand, we distribute information the same way but distribute uncertainty depending on the node degree according to \Cref{eq:correlation}, where $K^\mathrm{model}$ is a constant that regulates the scale of the uncertainty distribution, $\rho^\mathrm{model}$ is the correlation coefficient, $|| \mathbb{N}_i ||$ is the number of neighbors for agent $i$, and $N$ is the number of agents in the network. The correlation coefficient $\rho^\mathrm{model}$ lies between $-1$ and $1$ and shifts the uncertainty distribution from more central nodes being under-confident toward more central nodes being over-confident. We compare the performance for both types of errors across all networks as for the other experiments.

\begin{equation}
    \label{eq:correlation}
    {\sigma_0}^{[i]} = K^\mathrm{model} - \rho^\mathrm{model} 
 K^\mathrm{model} \frac{|| \mathbb{N}_i ||}{N}
\end{equation}

\subsection*{Different Methods for Opinion Updates}
\label{sec:opinion_methods}
We model the way each agent updates their opinion based on the opinion of their neighbors in multiple ways. The simplest way is direct averaging of the opinion of all neighbors, as used widely in the DeGroot model.\cite{degroot1974} We compare this naive averaging under different weighting schemes with two methods that in addition quantify uncertainty over the opinion and consider it when updating. Inspired by previous work in robotics, these methods locally approximate the uncertainty for each agent and use it to enable adaptive weighting of information, as we will further explain.
We will now first describe the naive averaging methods. Then we will explain further how we quantify uncertainty over opinion with Bayesian inference, relating the problem of opinion dynamics to a problem faced in robotics.

\subsubsection*{Naive Averaging Methods}
If we do not consider uncertainty over the opinion, we can directly update the opinion $x_{t+1}^{[i]}$ of each agent $i$ by averaging the opinion across its set of neighbors $\mathbb{N}_i$, as in \cref{eq:naive_update} using a weighting parameter $w_t^{[i]}$. The equation can be simplified to a weighted sum of all the neighbors, including the opinion of agent $i$. This form has an equivalent vectorial representation of the opinions, directly indicating the weights ${a_t}^{[i,j]}$ of the adjacency matrix of the network.
\begin{equation}\label{eq:naive_update}
    {x_{t+1}}^{[i]} = {w_t}^{[i]} {x_t}^{[i]} + \frac{1 - {w_t}^{[i]}}{|| \mathbb{N}_i ||} \sum_{j \in \mathbb{N}_i} {x_t}^{[j]} = \sum_{j \in \{\mathbb{N}_i, i\}} {a_t}^{[i,j]}{x_t}^{[j]}
\end{equation}

\paragraph*{Different Weighting Schemes}
To determine the weight parameter $w_t^{[i]}$, we can either set it universally for all agents or locally to reflect the number of neighbors. Universal weighting of the form $w_t^{[i]} = w_\mathrm{global}$ cannot reflect heterogeneity, while local weighting can easily be used to reflect the position of the individual within the network and thus heterogeneity of centrality. This can be achieved by putting equal weight on its own and each of its neighbors' opinions. Given the set of neighbors $\mathbb{N}_i$ of agent $i$, its weight thus needs to be set as $w_t^{[i]} = \frac{1}{|| \mathbb{N}_i || + 1}$. In our experiments, we compare such locally equal weights (NA-LEW) with the average of possible universal weights $w_\mathrm{global} \in [0, 0.1, \dots, 0.9, 1]$ (NA) and an optimal universal weight $w_\mathrm{global} = w_\mathrm{optimal}$ (NA-OUW). We obtain this optimal global weight by finding the best-performing weight concerning accuracy error for each parameter setting. Instead of choosing weights beforehand, we can estimate the uncertainty of agents to weight opinions \emph{adaptively} using Bayesian inference, as we now describe in further detail.

\subsubsection*{Bayesian Inference Methods}

When an agent updates its opinion, it combines multiple information sources: its last opinion and the opinions of its neighbors. However, these sources of information might be of different quality due to heterogeneity, as we described before. Thus, each agent needs to weigh these different sources differently to optimally combine them. A structurally similar problem is faced in robotics, where instead of interconnected heterogeneous agents, we have interconnected heterogeneous components. In these robotic systems, we estimate the uncertainty of each component with probabilistic modeling. We can then use Bayesian inference to combine the uncertain information within each component, leading to robust performance.\cite{martin2022coupled} We will apply the same principle here to update the opinion of each individual agent. To do so, we employ two different assumptions for Bayesian inference, leading to different opinion dynamics systems, as we will now elaborate.

\begin{figure}[t!]
    \centering
    \includegraphics[width=\textwidth]{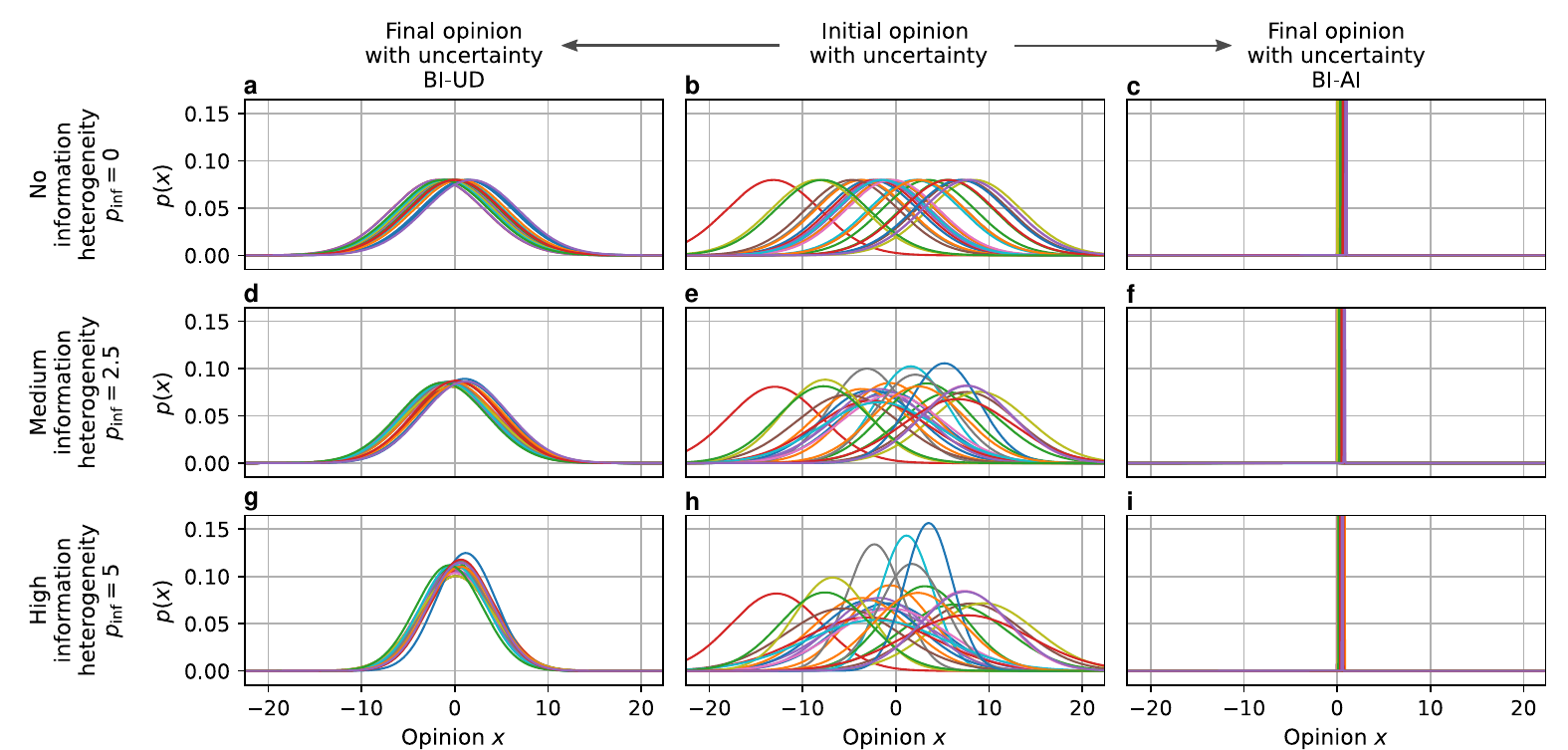}
    \caption{The gained certainty of agents varies with different inference assumptions and information heterogeneity. Assuming independence (BI-AI) agents obtain significantly high certainty over their opinions while assuming unknown dependence (BI-UD) leads only to a gain in certainty for the initially more uncertain agents.
    We show the initial opinions of 25 agents under varying levels of heterogeneity of initial information (the middle column) and final opinions following the two uncertainty-driven methods (left and right columns).
    As can be seen in the middle column, increasing information heterogeneity leads to a higher variation of uncertainty among agents. }
    \label{fig:het_inf_gen}
\end{figure}

\paragraph*{Bayesian Inference Assuming Independence}
For our case with initial Gaussian noise $\eta^{[i]}$, we can represent the belief over the opinion of each agent $i$ with a Gaussian $\mathcal{N}(\mu^{[i]},\,{\sigma^{2}}^{[i]})$ with mean ${\mu}^{[i]}$ and variance ${\sigma^2}^{[i]}$. After the initial distribution of information, this Gaussian reflects the initial opinion $x_0^{[i]}$ and its expected quality, which is quantified by the variance ${\sigma_0^2}^{[i]}$.

Before any exchange of information, these opinions and uncertainties of agents are independent of any other agent. Hence, we can use Bayes' theorem directly to update the belief over opinion $x^{[i]}$ of agent $i$ based on the belief over opinion of its neighbor $j$ (\cref{eq:bayes_basic}). Given the Gaussian belief representation, this means we update the two belief parameters for each agent based on the belief parameters of its neighbors. The inferred new mean ${\mu_{t+1}}^{[i]}$ corresponds to the average of its and its neighbors' opinions, weighted by their individual uncertainties (\cref{eq:bayes_update}), while the uncertainty over this new opinion, quantified as variance ${\sigma_{t+1}^2}^{[i]}$, is now reduced according to the uncertainties of all neighbors following \cref{eq:bayes_unc_update}. Due to the assumed independence, the uncertainty of each agent monotonically decreases, quickly leading to highly certain opinions as shown in \Cref{fig:het_inf_gen}.

\begin{align}
    \label{eq:bayes_basic}
    p(x^{[i]} | x^{[j]}) &= \frac{p(x^{[j]} | x^{[i]}) p(x^{[i]})}{p(x^{[j]})}\\
    \label{eq:bayes_update}
    {\mu_{t+1}}^{[i]} &= \Big(\frac{{\mu_t}^{[i]}}{{\sigma_t^2}^{[i]}} + \sum_{j \in \mathbb{N}_i} \frac{{\mu_t}^{[j]}}{{\sigma_t^2}^{[j]}}\Big) \cdot \frac{1}{\frac{1}{{\sigma_t^2}^{[i]}} + \sum_{j \in \mathbb{N}_i} \frac{1}{{\sigma_t^2}^{[j]}}}\\
    \label{eq:bayes_unc_update}
    {\sigma_{t+1}^2}^{[i]} &= \frac{1}{\frac{1}{{\sigma_t^2}^{[i]}} + \sum_{j \in \mathbb{N}_i} \frac{1}{{\sigma_t^2}^{[j]}}}
\end{align}

In the results, we call this updating of opinion and uncertainty with Bayesian inference assuming independence (BI-AI). It corresponds to the inverse-variance weighted average of all opinions within the neighborhood or a one-dimensional Kalman filter\cite{kalman1960new} where each neighbor's opinion represents an independent measurement. Thus, the overall collective represents a set of interconnected Kalman filters, similar to the set of interconnected extended Kalman filters used in the related robotics work.\cite{martin2022coupled}

After just one round of interactions between agents, their beliefs are, however, no longer independent. Moreover, with increasing exchanges, their dependence rises. Nonetheless, we will employ the independence assumption as approximation, because it does not just offer tractability but as we show also interesting analytical and performance benefits compared to a more conservative approximation of uncertainty, which we will now describe further.

\paragraph*{Bayesian Inference Assuming Unknown Correlation}
Because the independence assumption does not hold in collective opinion dynamics, we need to take possible correlation of opinions into account. But as we already explained, tracking all correlations in the collective is an NP-hard problem.\cite{hazla_bayesian_2021,chickering_large-sample_nodate} Instead of following this intractable approach, we thus assume unknown correlation of opinions and update the beliefs of each agent accordingly without tracking any correlation.

Assuming unknown correlation, the optimal new mean still follows \cref{eq:bayes_update} but the best possible conservative estimate of its variance becomes a similarly weighted average as well:
\begin{equation}\label{eq:ci_update}
        {\sigma_{t+1}^2}^{[i]} = \frac{1}{\Big(\frac{1}{{\sigma_t^2}^{[i]} {\sigma_t^2}^{[i]}} + \sum_{j \in \mathbb{N}_i} \frac{1} {{\sigma_t^2}^{[j]} {\sigma_t^2}^{[j]}}\Big) \cdot \frac{1}{\frac{1}{{\sigma_t^2}^{[i]}} + \sum_{j \in \mathbb{N}_i} \frac{1}{{\sigma_t^2}^{[j]}}}} = \frac{\frac{1}{{\sigma_t^2}^{[i]}} + \sum_{j \in \mathbb{N}_i} \frac{1}{{\sigma_t^2}^{[j]}}}{\frac{1}{({\sigma_t^2}^{[i]})^2} + \sum_{j \in \mathbb{N}_i} \frac{1}{({\sigma_t^2}{[j]})^2}} \quad .
\end{equation}

In the results, we call this updating rule Bayesian inference assuming unknown dependence (BI-UD). It leads to each update resembling one-dimensional covariance intersection of all the measurements,\cite{julier1997non} directly implementing its fast variant.\cite{niehsen2002information} Because of the conservative uncertainty updates, the agents remain more uncertain while their opinions converge, as we show in \Cref{fig:het_inf_gen}.

\subsection*{Metrics}

\subsubsection*{Accuracy}
We evaluate the collective performance according to the expected value of agents' opinions $\mathbb{E}[x^{[i]}]$. This way, we have a generic metric for both the opinions having a distribution, such as in~\Cref{eq:bayes_basic}, and the scalar ones, as in the DeGroot model. Similar to our previous work,~\cite{raoufi2021speed} we decompose the total accuracy error ($E^\text{A}$) into precision and trueness errors. Precision error ($E^\text{P}$) measures the variance of opinions based on which we measure and define the convergence of opinions to the collective average (${x_{t}}^{\text{col}}=\sum_{i=1}^{N} {\mathbb{E} [ {x_{t}}^{[i]} ] } / {N}$). The bias of the opinions compared to the neutral opinion is quantified by the trueness error ($E^\text{T}$). Interpreting opinions as individual estimation errors with the zero-mean initial distribution, we assumed this neutral opinion to be zero (${x_{t}}^{\text{neutral}}=0$). We define the metrics as

\begin{align}
    {E_{t}}^\text{P} =& \frac{1}{\text{N}} \sum\limits_{i=1}^\text{N} \left ( {\mathbb{E} \left [x_{t}^{[i]} \right ] } - x_{t}^{\text{col}}  \right )^2 \ , \quad
    {E_{t}}^\text{T} = \left ( x_{t}^{\text{col}} - {x_{t}^{\text{neutral}} } \right ) ^2 \ , \quad
    {E_{t}}^\text{A} = \frac{1}{\text{N}} \sum\limits_{i=1}^\text{N} \left ( {\mathbb{E} \left [x_{t}^{[i]} \right ] } - {x_{t}}^{\text{neutral}} \right ) ^2 = {E_{t}}^\text{T} + {E_{t}}^\text{P} \ . 
\end{align}

Given that we used a fixed duration for the simulation and the same initial distributions, a lower precision error at the end of simulations indicates a higher speed of convergence.

\subsubsection*{Centrality of Nodes}

Among different definitions of node centrality, we used the node in-degree as its current centrality metric. We assume the weighted adjacency matrix of the network ($\textbf{A}=(a^{[i,j]})$) is row-normalized, i.e., for each row $i$ the sum of elements equals one ($ \sum_{j} a^{[i,j]} = 1 $). This means that the out-degree of each node is equal to one, ensuring that the opinions of agents do not diverge. The in-degree of the node $i$ is calculated as the sum of $i$-th column of the adjacency matrix, $d_i =  \sum_{j=1}^{N} a^{[j,i]}$ .

\subsubsection*{Heterogeneity of Centrality}
By taking the node in-degree as a measure of the node centrality, we use the standard deviation of the degree distribution as the indicator of heterogeneity in the centralization of the \textit{unweighted} graph.~\cite{snijders1981degree, jacob2017measure} Here, a greater standard deviation means that the difference between the in-degree of the most central and peripheral nodes is higher, suggesting a higher heterogeneity of centrality. We quantify $\sigma_\text{net}$ as:
\begin{equation}
\sigma_\text{net}(d) = \sqrt{\frac{1}{N} \sum_{i=1}^{N} (d_i - \langle d \rangle)^2}, 
\end{equation}
where $d_i$ is the $i$-th node's in-degree of the graph $G$ with the mean degree $\langle d \rangle$.

\subsection*{Simulation Configuration}
With the above-described methods for network generation, information distribution, and opinion updating, we simulate opinion dynamics in heterogeneous collectives. For these experiments, we choose $\text{N} = 100$ agents and simulate them for $T=10$ time steps at which point they converge toward one dynamic of reducing precision error at a constant trueness error, as can be seen in \Cref{fig:overtime}. Updates of all agents occur simultaneously and do not depend on the order of agents and their updates.

To ensure a good approximation of the different dynamics, we simulate $100$ Monte Carlo samples for each configuration of parameters and report the average over these samples. This is except for the qualitative results in \Cref{fig:overtime,fig:kreg_uncert_central}. For the heterogeneity parameters, $p_\mathrm{inf}$, $p_\mathrm{inf}^\mathrm{model}$, and $p_r$, we sample parameter settings on a logarithmic scale to obtain more information around the shifting point of the dynamics. This means for heterogeneity of information, $p_\mathrm{inf}$ and $p_\mathrm{inf}^\mathrm{model}$, obtaining more samples close to $0$ where there would be absolute information homogeneity and obtaining more samples close to $0$ for heterogeneity of centrality $p_r$, where the shift from a loop-free star network occurs. For other parameters, we choose them as described in the previous sections, sampling them on a linear scale on their permissible interval.

\bibliography{p27_p35}

\section*{Data Availability}
The datasets generated and analyzed during the current study \textit{will be} made available online. %
\section*{Code Availability}
The underlying code for this study \textit{will be} available on GitHub and can be accessed via this link \url{https://github.com/mohsen-raoufi/uncertainty_opinion_dynamics}. 

\section*{Acknowledgements}

Funded by the Deutsche Forschungsgemeinschaft (DFG, German Research Foundation) under Germany’s Excellence Strategy – EXC 2002/1 “Science of Intelligence” – project number 390523135.

\section*{Author contributions statement}
V.M. and M.R. contributed equally to conceiving and implementing models and experiments, as well as to analyzing the results. O.B., H.H., and P.R. equally supervised the project. V.M. and M.R. wrote the manuscript with input from all authors. All authors reviewed the manuscript.

\section*{Additional information}
The authors declare no competing interests.

\end{document}